\documentstyle[aps,prb,multicol,epsf]{revtex}
\begin{document} 
\draft
\title{The $E\otimes e$ Jahn-Teller Polaron 
in Comparison with the Holstein Polaron}
\author{Yasutami Takada}

\address{Institute for Solid State Physics, University of Tokyo, 
7-22-1 Roppongi, Minato-ku, Tokyo 106-8666, Japan}
\date{\today}
\maketitle
\begin{abstract}
Based on an exact expression for the self-energy of the Jahn-Teller 
polaron, we find that symmetry of pseudospin rotation makes the 
vertex correction much less effective than that for the Holstein 
polaron. 
This ineffectiveness brings about a smaller effective mass $m^*$ 
and a quantitatively differenent large-to-small polaron crossover, 
as examined by exact diagonalization in a two-site system. 
In the strong-coupling and antiadiabatic region, a rigorous analytic 
expression is found for $m^*$. 
\end{abstract}


\begin{multicols}{2}
\narrowtext
It is well recognized that both the double-exchange mechanism and 
the strong electron-phonon interaction, specifically the Jahn-Teller 
(JT) effect on doubly-degenerate $e_g$ orbitals coupled with two 
degenerate vibrations (the $E\otimes e$ case) at each Mn$^{3+}$ 
site, are essential ingredients to bring about the colossal 
magnetoresistance (CMR) in manganese-oxide 
perovskites.\cite{Millis1,Muller,Billinge,Mareo} 
Thus the theories on CMR need to include these ingredients 
simultaneously. \cite{Bishop,Millis2,Min}
This complicated situation compels some theories to neglect kinetic 
energies of ions and others to treat the JT polaron in a way similar 
to the conventional polaron. \cite{Frohlich,Holstein} 
In either way, characteristic features of the JT polaron 
do not emerge from those theories. 

In fact, in spite of a broad interest in its role in 
superconductivity, studies on the JT effect in an itinerant electron 
system are limited, probably because the word ``the JT effect'' 
often implies strong lattice deformations and a localized electron 
associated with them. 
H\"ock {\it et al.} \cite{Hock} considered the simplest case, 
namely, the $E\otimes \beta$ JT polaron which, unfortunately, 
possesses a too simple internal structure to provide qualitatively 
different features from those of the Holstein 
polaron.\cite{Holstein} 
The second simplest $E\otimes e$ case was treated by Fabrizio and 
Tosatti \cite{Fabrizio} as well as 
Benedetti and Zeyher,\cite{Benedetti} 
but both works addressed only localization 
in the strong-coupling region. 

In this paper, we provide a new aspect to this problem by 
making a comparative study of the $E\otimes e$ JT polaron with 
the Holstein polaron based on the knowledge attained after 
forty-year's investigation into the latter.\cite{Alexandrov} 

We have obtained the following results for the system specified 
by the two parameters, $\tilde{t} \! \equiv \! t/\omega_0$ and 
$\alpha \! \equiv \! E_{\rm JT}/\omega_0$, where $t$, $\omega_0$, 
and $E_{\rm JT}$ are the energies corresponding to bare electron 
transfer, bare phonon, and Jahn-Teller stabilization, respectively. 
(1) Based on an expression for the self-energy derived by a similar 
method for the Fr\"ohlich polaron,\cite{Whitefield} 
we find that the vertex correction is much less effective in the 
JT polaron than that in the conventional polarons due to a local 
conservation law imposed on the JT Hamiltonian.\cite{Bersuker} 
(2) This ineffectiveness leads us to a smaller effective mass, 
as shown by an explicit expression for the JT polaron-mass 
enhancement factor as $\sqrt{2/\pi \alpha}\,e^{\alpha}$ 
in the antiadiabatic ($\tilde{t} \! \ll \! 1$) and strong-coupling 
($\alpha \! \gg \! 1$) region. 
(3) The large-to-small polaron crossover is examined by exact 
diagonalization (ED) in a two-site system on the ground that 
the ED calculation on small clusters is very effective for 
$\alpha \! \gg \! 1$. 
We find that the crossover occurs at $\tilde{t} \! \approx \! \alpha 
\!-\! \sqrt{\alpha}$, indicating that JT polarons are more mobile 
than Holstein ones. 

Let us start with a single $E\otimes e$ center at site ${\bf j}$ 
described by the Hamiltonian $H_{\bf j}$ as \cite{spin}
\begin{equation}
 \label{site-Hamiltonian}
 H_{\bf j} \!= \!A 
\bigl [ q_{{\bf j}a} (\!d^{+}_{{\bf j}a}d_{{\bf j}b}
\!+\! d^{+}_{{\bf j}b}d_{{\bf j}a}\!) 
\!+\!q_{{\bf j}b} (\!d^{+}_{{\bf j}a}d_{{\bf j}a}
\!-\! d^{+}_{{\bf j}b}d_{{\bf j}b}\!) \bigr ] \!
\!+\! H_{{\bf j}{\rm v}}, 
\end{equation}
with $A\!=\!\omega_0\sqrt{2E_{\rm JT}}$, where $d_{{\bf j}a}$ and 
$d_{{\bf j}b}$ represent electron annihilation operators 
for the two degenerate orbitals, $q_{{\bf j}a}$ and $q_{{\bf j}b}$ 
are the two local JT distortions, and $H_{{\bf j}{\rm v}}$ is 
the harmonic Hamiltonian for the vibrational modes. 
In polar coordinates as 
$q_{{\bf j}a}\! =\! q_{\bf j} \sin \theta_{\bf j}$ and 
$q_{{\bf j}b}\! =\! q_{\bf j} \cos \theta_{\bf j}$, 
the energy eigenfunction for $H_{{\bf j}{\rm v}}$, 
$\langle q_{\bf j}\theta_{\bf j} | nl \rangle$, satisfying 
$H_{{\bf j}{\rm v}} | nl \rangle\!=\! \omega_0 (n+1) | nl \rangle$
is given by \cite{ion-mass}
\begin{equation}
 \label{phonon-eigenfunction}
\langle q_{\bf j}\theta_{\bf j} | nl \rangle
 =   N_{l,p}
F(-p,|l|\!+\!1,z_{\bf j})z_{\bf j}^{|l|/2}
e^{-z_{\bf j}/2} e^{il\theta_{\bf j}},
\end{equation}
with $N_{l,p}\!=\!(-1)^{(|l|\!-\!l)/2}
\sqrt{\omega_0(|l|\!+\!p)!/\pi p!}/|l|!$, 
$z_{\bf j} \! \equiv\! \omega_0 q_{\bf j}^2$, and 
$n\!=\!|l|\!+\!2p$, where $F(-p,|l|\!+\!1,z_{\bf j})$ 
is the confluent hypergeometric function, $l$ is an integer, 
and $p\!=\!0,1,2,\cdots$. 

In terms of boson operators, $a_{\bf j}$ and $b_{\bf j}$, to 
represent $q_{{\bf j}a}$ and $q_{{\bf j}b}$ in second 
quantization, $H_{\bf j}$ is rewritten as 
\begin{eqnarray}
 \label{site-Hamiltonian2}
 H_{\bf j} &=& \omega_0 \sqrt{2\alpha} 
\bigl [ (a^{+}_{\bf j}\!-\!b_{\bf j})c^{+}_{{\bf j}\uparrow} 
c_{{\bf j}\downarrow} 
\!+\! (a_{\bf j}\!-\!b^{+}_{\bf j})c^{+}_{{\bf j}\downarrow} 
c_{{\bf j}\uparrow} \bigr ]
\nonumber \\
&&+\omega_0 (a^{+}_{\bf j}a_{\bf j}\!+\!
b^{+}_{\bf j}b_{\bf j}\!+\!1),
\end{eqnarray}
where pseudospin index $\sigma(=\! \pm 1\! =\,\uparrow {\rm or} 
\downarrow)$ for electron operators is introduced through the 
relation $c_{{\bf j}\sigma} \!\equiv\! (d_{{\bf j}a}\!+\!
i\sigma d_{{\bf j}b})/\sqrt{2}$. 
Note that second-quantized representation for phonons is not 
unique due to $SU(2)$ symmetry in $H_{{\bf j}{\rm v}}$. 
We have chosen it in such a way as to diagonalize both 
$H_{{\bf j}{\rm v}}$ and $\hat{l}_{\bf j} \! \equiv \! -i\partial
/\partial \theta_{\bf j}$. 
Then we obain $| nl \rangle$ as
\begin{equation}
 \label{phonon-eigenfunction2}
| nl \rangle \! = \! 
{1 \over \sqrt{[(n\!+\!l)/2]![(n\!-\!l)/2]!}}
{a^{+}_{\bf j}}^{n+l \over 2}
{b^{+}_{\bf j}}^{n-l \over 2} \,
| {\rm vacuum} \rangle .
\end{equation}
This phonon representation is a key step to obtain analytic 
expressions in Eqs.~(\ref{ground-state}) and (\ref{effective-mass}) 
as well as a clear view of less effectiveness 
of the vertex correction. 

Because of the symmetry of pseudospin rotation, the operator 
$L_{\bf j}$, defined by $L_{\bf j} \equiv 
a^{+}_{\bf j}a_{\bf j}\!-\!b^{+}_{\bf j}b_{\bf j}\! - \! 
(c^{+}_{{\bf j}\uparrow} c_{{\bf j}\uparrow}
\! - \! c^{+}_{{\bf j}\downarrow} c_{{\bf j}\downarrow})/2$, 
is conserved as easily checked by 
$[H_{\bf j},L_{\bf j}]\!=\!0$. 

For a single electron at site ${\bf j}$, eigenvalues for 
$L_{\bf j}$ are half-integers and each energy level is doubly 
degenerate corresponding to $\pm |L_{\bf j}|$. 
In general we can give the ground-state wavefunction 
$\Psi^{(0)}_{\bf j}$ only numerically, but for large $\alpha$ 
we find an analytic expression as 
\begin{equation}
 \label{ground-state}
\Psi^{(0)}_{\bf j} \approx 
{\sqrt{2/\alpha}\ b_{\rm j}c^{+}_{{\bf j}\uparrow}
\! + \! c^{+}_{{\bf j}\downarrow} \over 
\sqrt{I_0(\alpha)\!+\!I_1(\alpha)}}
J_0\Bigl (\sqrt{2\alpha a^{+}_{\bf j} b^{+}_{\bf j}}\,\Bigr )
| {\rm vacuum} \rangle ,
\end{equation}
for $L_{\bf j}\!=\!1/2$, \cite{negative} 
where $J_0(x)$ is the Bessel function and $I_i(x)$ 
its modified form. 
The corresponding energy $E_0$ is given as 
$E_0 \approx (-\alpha +1/2+ 1/16\alpha)\omega_0 
\approx -E_{\rm JT}$. 

Now we consider a lattice composed of $N$ JT centers for which 
the Hamiltonian is given as $H_{\rm JT}=H_t+\sum_{\bf j}H_{\bf j}$, 
where $H_t$ describes the transfer energies between nearest-neighbor 
JT centers as
\begin{equation}
 \label{transfer-Hamiltonian}
 H_t = -\sum_{<{\bf jj'}>}\sum_{\gamma=a}^b\sum_{\gamma'=a}^b
t_{\gamma \gamma'}(d^{+}_{{\bf j}\gamma}d_{{\bf j'}\gamma'} + 
d^{+}_{{\bf j'}\gamma'}d_{{\bf j}\gamma}). 
\end{equation}
For simplicity, we take $t_{\gamma \gamma'}\!=\!
\delta_{\gamma \gamma'}t$ in the following. 
Then Eq.~(\ref{transfer-Hamiltonian}) can be rewritten as
\begin{equation}
 \label{transfer-Hamiltonian2}
 H_t = -t \sum_{<{\bf jj'}>}\sum_{\sigma}
(c^{+}_{{\bf j}\sigma}c_{{\bf j'}\sigma} + 
c^{+}_{{\bf j'}\sigma}c_{{\bf j}\sigma})
= \sum_{{\bf k}\sigma}\varepsilon_{\bf k}
c^{+}_{{\bf k}\sigma}c_{{\bf k}\sigma},
\end{equation}
where $c_{{\bf k}\sigma}(=\! N^{-1/2}\sum_{\bf j}
e^{-i{\bf j}\cdot {\bf k}}c_{{\bf j}\sigma})$ 
is the Fourier transform of $c_{{\bf j}\sigma}$ and 
$\varepsilon_{\bf k}$ represents its bare dispersion relation. 
Note that the operator $L$ defined by $L \! \equiv \! \sum_{\bf j} 
L_{\bf j}$ is conserved, namely, $[H_{\rm JT},L]\!=\!0$ in this 
choice of $t_{\gamma \gamma'}$. 

The thermal one-electron Green's function 
$G_{{\bf k}\sigma}(i\omega_n)$ with $\omega_n$ a fermion Matsubara 
frequency is defined by \cite{units} 
\begin{equation}
 \label{green's-function}
G_{{\bf k}\sigma}(i\omega_n) 
= \int_0^{\beta}d\tau e^{i\omega_n \tau} G_{{\bf k}\sigma}(\tau), 
\end{equation}
with $\beta=T^{-1}$ and 
$G_{{\bf k}\sigma}(\tau)\!\equiv\!-\langle T_{\tau}
c_{{\bf k}\sigma}(\tau)c^{+}_{{\bf k}\sigma}\rangle$. 
We first consider $\partial G_{{\bf k}\sigma}(\tau)/\partial \tau$ 
to derive an equation of motion which relates 
$G_{{\bf k}\sigma}(\tau)$ with the electron-phonon correlation 
function $\langle T_{\tau}\sum_{\bf q}\{
[a_{\bf q}^{+}(\tau)\!-\!b_{\bf -q}(\tau)]
c_{{\bf k+q}-\sigma}(\tau)c_{{\bf k}\sigma}^{+}\}\rangle$. 
Next we derive a similar equation of motion for this 
correlation function in order to eliminate phonon operators 
in the expressions other than the bare phonon Green's 
function which is the same for both phonons as 
$D_0(i\omega_{m}) \!=\! 2\omega_0/
[(i\omega_{m})^2\!-\!{\omega_0}^2]$ with $\omega_{m}$ a boson 
Matsubara frequency. 
Then we arrive at an exact expression for the self-energy 
$\Sigma_{{\bf k}\sigma}(i\omega_n)$ as 
\begin{eqnarray}
 \label{self-energy}
\Sigma_{{\bf k}\sigma}(i\omega_n)&=& 
-T\sum_{\omega_{n'}}\sum_{\bf k'}{2\alpha \over N}{\omega_0}^2 
D_0(i\omega_{n'}\!-\!i\omega_n)
\nonumber \\
&& \times G_{{\bf k'}-\sigma}(i\omega_{n'}) 
\Lambda_{-\sigma \sigma}({\bf k'},i\omega_{n'};
{\bf k},i\omega_{n}),
\end{eqnarray}
where the vertex function 
$\Lambda_{\sigma' \sigma}({\bf k'},i\omega_{n'};
{\bf k},i\omega_{n})$, a key quantity in this expression, 
is found to be 
\begin{eqnarray}
 \label{vertex-function}
G_{{\bf k}\sigma}&&(i\omega_n)G_{{\bf k'}\sigma'}(i\omega_{n'})
 \Lambda_{\sigma' \sigma}({\bf k'},i\omega_{n'};
{\bf k},i\omega_{n}) \!= \! \int_0^{\beta}\! d\tau 
\,e^{i\omega_{n'}\tau}
\nonumber \\
&&\times \int_0^{\beta}\! d\tau' \, 
e^{i(\omega_{n}-\omega_{n'})\tau'} \!
\langle T_{\tau}c_{{\bf k'}\sigma'}(\tau)
S^{\sigma'\sigma}_{{\bf k'}\!-\!{\bf k}}(\tau')
c^{+}_{{\bf k}\sigma} \rangle, 
\end{eqnarray}
with $S^{\sigma'\sigma}_{{\bf k'}\!-\!{\bf k}} \! \equiv \! 
\sum_{\bf k''}
c^{+}_{{\bf k''}\!+\!{\bf k'}\!-\!{\bf k}\sigma'}
c_{{\bf k''}\sigma}$, reflecting the spinor nature of the problem. 
Equation~(\ref{self-energy}) serves as a firm basis to study 
the JT polaron in the Green's function approach. 

Quite an analogous result has been obtained for the conventional 
polaron. \cite{Whitefield}
For the Holstein model specified by the Hamiltonian $H_{\rm H}$ as 
\begin{eqnarray}
 \label{Holstein-Hamiltonian}
H_{\rm H} \!&=&\! \sum_{{\bf k}\sigma}\varepsilon_{\bf k}
c^{+}_{{\bf k}\sigma}c_{{\bf k}\sigma} \!+\! 
\omega_0 \sqrt{2\alpha} \sum_{{\rm j}\sigma} 
(a^{+}_{\bf j}\!+\!a_{\bf j})c^{+}_{{\bf j}\sigma} 
c_{{\bf j}\sigma} 
\nonumber \\
&& + \omega_0 \sum_{\bf j}(a^{+}_{\bf j}a_{\bf j}\!+\! 1/2),
\end{eqnarray}
where $\sigma$ in this case refers to ``real'' spin index, 
$\Sigma_{{\bf k}\sigma}(i\omega_n)$ is given 
in the form of Eq.~(\ref{self-energy}) in which 
$G_{{\bf k'}-\sigma}(i\omega_{n'})$ and 
$\Lambda_{-\sigma \sigma}({\bf k'},i\omega_{n'};
{\bf k},i\omega_{n})$ are, respectively, changed into 
$G_{{\bf k'}\sigma}(i\omega_{n'})$ and 
$\Lambda_{c}({\bf k'},i\omega_{n'};{\bf k},i\omega_{n})$ 
the charge vertex function, defined through 
Eq.~(\ref{vertex-function}) with 
$S^{\sigma'\sigma}_{{\bf k'}\!-\!{\bf k}}$ replaced by 
the charge operator $\rho_{{\bf k'}\!-\!{\bf k}} \! \equiv \! 
\sum_{\bf k''\sigma''}
c^{+}_{{\bf k''}\!+\!{\bf k'}\!-\!{\bf k}\sigma''}
c_{{\bf k''}\sigma''}$ due to the scalar nature 
of the Holstein system. 

The diagram to represent Eq.~(\ref{self-energy}) is given 
in Fig.~1(a), in which we introduce the vertex $\Gamma$ by 
eliminating improper diagrams from the vertex $\Lambda$. 
The expansion series for $\Gamma$ in terms of $\alpha$ is shown 
in Fig.~1(b). 
If we assume that $G_{{\bf k}\sigma}(i\omega_n)$ is independent of 
$\sigma$ and emply the Migdal's approximation \cite{Migdal} 
in which only $\Gamma_0$ is retained for $\Gamma$, namely, 
$\Gamma\!=\! 1$, there exists no difference in the self-energy 
between JT and Holstein systems. 
\begin{figure}[h]
\centerline{\epsfxsize=2.4truein \epsfbox{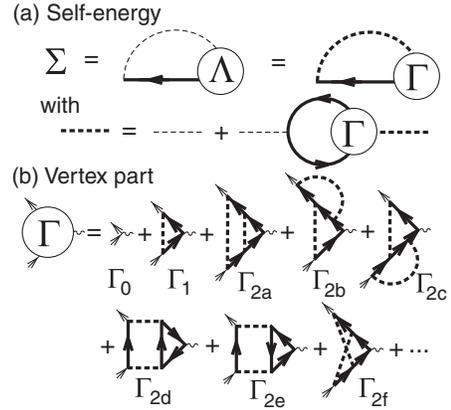} }
\caption{(a) Self-energy in diagrammatic representation. 
Thick solid and thin dashed lines indicate, respectively, 
the electron Green's function and the bare phonon propagator. 
(b) Expansion series for the vertex $\Gamma$ up to second order 
in $\alpha$.} 
\end{figure}

There is, however, an important difference in the vertex correction. 
In contrast to the Holstein system, the corrections represented by 
the diagrams $\Gamma_1, \Gamma_{2a}, \cdots, \Gamma_{2e}$ is seen 
to vanish in the JT system by merely considering the pseudospin 
assignment together with the direction of phonon propagators, 
because Eq.~(\ref{site-Hamiltonian2}) dictates that the JT-phonon 
exchange interaction works only in the pseudospin exchange process 
between electrons with opposite pseudospins. 
Physically both electrons and phonons in the JT system are associated 
with a notion of clockwise- or counterclockwise-``rotation'' around 
each JT center and electrons interact with phonons only when 
the total rotation is conserved. 
In this sense, the vanishment of these vertex corrections is due to 
the local-rotation conservation law. 
This law allows only processes such as the one represented by 
$\Gamma_{2f}$ for $\Gamma$. 
Similarly, all the third-order vertex corrections vanish. 
Ineffectiveness of the vertex correction widens the applicable 
range in $\alpha$ of the Migdal's approximation in the JT system 
and it leads us to the smaller polaron mass enhancement factor 
$m^*/m$ than that in the Holstein model in which the correction 
$\Gamma_1$ is known to enhances $m^*/m$ as $\alpha$ increases. 

The above perturbative approach is not useful in discussing a small 
polaron or polaron localization in a site. 
According to the studies on the Holstein model, \cite{Alexandrov} 
an ED calculation in a two-site system provides qualitatively correct 
and quantitatively fair results for the small polaron in the 
strong-coupling region ($\alpha \! > \! 1$), irrespective of the 
value of $\tilde{t}$. 
Thus we shall make a similar analysis of a single electron in the JT 
model with $N=2$ in which the eigenvalues of the conserved quantity 
$L$ are half-integers and each energy level is doubly degenerate. 

Let us consider the antiadiabatic region 
($\tilde{t} \! \ll \! 1$) first. 
At $\alpha \! \gg \! 1$, both the ground and first-excited states 
belong to the sector of $|L|\! = \! 1/2$. 
Using $\Psi^{(0)}_{j}$ in Eq.~(\ref{ground-state}), 
their wavefunctions $\Psi_{\pm}$ are written as 
$\Psi_{\pm}\! \approx \! 
(\Psi^{(0)}_{1} \pm \Psi^{(0)}_{2})/\sqrt{2}$ for $L\!=\!1/2$ 
with the corresponding energies $E_{\pm}\! =\! 
E_0 \pm t/[I_0(\alpha)\!+\!I_1(\alpha)]$. 
The energy difference, $E_{+}-E_{-}$, can be used to estimate the 
polaron bandwidth in a crystal and thus its ratio with the bare value, 
$2t$, determines the polaron effective mass through the relation 
\begin{equation}
 \label{effective-mass}
{m \over m^*} = {E_{+}-E_{-} \over 2t} 
= {1 \over I_0(\alpha)\!+\!I_1(\alpha)} \approx 
\sqrt{{\pi \alpha \over 2}}\,e^{-\alpha}. 
\end{equation}
This result should be compared with $e^{-2\alpha}$ 
the Holstein's famous result \cite{Holstein} for the system defined 
in Eq.~(\ref{Holstein-Hamiltonian}). 

We resort to ED calculations to obtain $m/m^*$ through the numerical 
evaluation of $E_{\pm}$ for arbitrary $\alpha$. 
The conservation of $L$ helps reduce the number of expansion 
bases for phonons considerably. 
We plot the calculated $m/m^*$ for both JT and Holstein models in 
Fig.~2 in which $\tilde{t}$ is taken as $0.2$, although the result 
itself does not depend on $\tilde{t}$ provided that it is much smaller 
than unity. 
(The result for the Holstein model hardly changes from the analytic 
result $e^{-2\alpha}$ in the whole region of $\alpha$.) 
For small $\alpha$, both models give essentially the same 
$m/m^*$ as implied by the previous weak-coupling analysis. 
For large $\alpha$, however, there is a difference in $m^*/m$ which 
is more than orders of magnitude for $\alpha\! >\! 1$, indicating 
that the JT polaron is quite mobile compared to the Holstein polaron. 
\begin{figure}[h]
\centerline{\epsfxsize=2.6truein \epsfbox{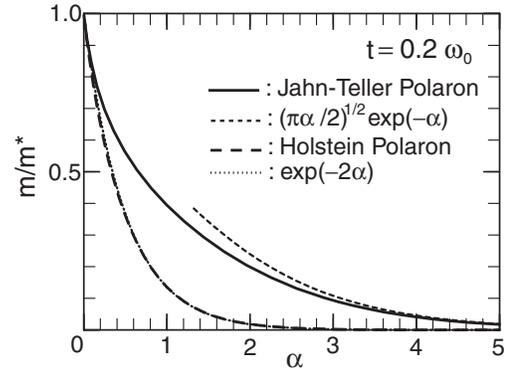} }
\caption{Polaron mass reduction factor $m/m^*$ for the JT (the solid 
curve) and the Holstein (the dashed curve) models with each analytic 
expression in the strong-coupling region.} 
\end{figure}

Next we make a semiclassical argument on the adiabatic region 
($\tilde{t} \! \gg \! 1$) by considering the adiabatic potential 
$U_{\rm ad}$ for given phonon variables, 
$\{q_{1}\theta_{1};q_{2}\theta_{2}\}$. 
Since it was calculated previously in connection with the 
Berry phase,\cite{Takada} we just give the result here as 
\begin{eqnarray}
 \label{adiabatic-potential}
U_{\rm ad} \! &=& \! 
{\omega_0^2 \over 2}q^2 
\! -\! \Bigl \{ t^2\!  +\!  \alpha {\omega_0}^3 q^2 
\!+\! \bigl \{ 2\alpha {\omega_0}^3 t^2 
[q^2\!+\! 2q_1q_2
\nonumber \\
&& \times \cos (\theta_1\! -\! \theta_2)]\!+\!
\alpha^2 {\omega_0}^4 
 ({q_{1}}^2\! -\! {q_{2}}^2)^2 \bigr \}^{1/2}\Bigr \}^{1/2}, 
\end{eqnarray}
with $q^2 \! \equiv \! {q_{1}}^2\! +\! {q_{2}}^2$. 
This potential has rather simple features; if the adiabaticity 
parameter $\lambda \! \equiv \! \alpha/\tilde{t}\! =\! 
E_{\rm JT}/t$ is less than unity, $U_{\rm ad}$ has only one 
minimum in $\{q_{1}\theta_{1};q_{2}\theta_{2}\}$-coordinate space, 
implying no symptom for a small polaron. 
On the other hand, it is a double-well potential for 
$\lambda \! > \! 1$ with the energy barrier 
$\Delta \! =\! (\alpha \omega_0/2) (1\! -\! \lambda^{-1})^2$. 
If the largest zero-point energy of phonons $\Delta_{\rm zero}$ 
(which is $\omega_0/2$ in this case) is smaller than $\Delta$, 
localization leading to a small polaron occurs. 
Thus the condition $\Delta \! \gtrsim \! \Delta_{\rm zero}$ provides 
the criterion to obtain a small polaron as 
\begin{equation}
 \label{small-polaron}
\tilde{t} \lesssim \alpha - \sqrt{\alpha}, \quad {\rm with}\ 
\alpha > 1. 
\end{equation}
A similar argument has been done for the Holstein model described 
in Eq.~(\ref{Holstein-Hamiltonian}) for which a double-well 
potential appears only when $\lambda \! > \! 1/2$ with 
$\Delta \! = \! \alpha \omega_0 (1\!-\! 1/2\lambda)^2$ and 
$\Delta_{\rm zero} \! = \! (\omega_0/2)\sqrt{1\! -\! 1/4\lambda^2}$ 
\cite{Alexandrov}, leading to the criterion 
\begin{equation}
 \label{small-polaron2}
\alpha \Bigl (1-{1 \over 2\lambda}\Bigr )^2 \gtrsim 
{1 \over 2} \sqrt{1 -{1 \over 4\lambda^2}}\,. 
\end{equation}
This condition cannot be reduced to such a simple form as that 
in (\ref{small-polaron}), but clearly it is much less restrictive 
than (\ref{small-polaron}) for the small-polaron formation. 

Finally we make a more quantitative argument on the large-to-small 
polaron crossover based on the exact ground-state wavefunction 
$\Psi_0$ obtained by the ED calculation. 
We evaluate two quantitites, ``the transfer amplitude per bond'' 
$T\! \equiv \! \langle \Psi_0 | \sum_{\sigma}
(c^{+}_{1\sigma}c_{2\sigma}\! + \!c^{+}_{2\sigma}c_{1\sigma})
|\Psi_0 \rangle/N_{\rm bond}$ with the number of the bond 
$N_{\rm bond}\!=\!1$ and ``the interaction amplitude per site'' 
$I \! \equiv \!  \langle \Psi_0 | \sum_{j} \bigl [
(a^{+}_{j}\!-\!b_{j})c^{+}_{j\uparrow} c_{j\downarrow} 
\!+\! (a_{j}\!-\!b^{+}_{j})c^{+}_{j\downarrow} c_{j\uparrow} 
\bigr ]|\Psi_0 \rangle /N$ with $N\!=\!2$. 
Then we measure ``itineracy'' by the ratio $|T/I|$, 
because the ratio must be large for an itinerant polaron. 

Contour plots for $|T/I|$ in $(\tilde{t},\alpha)$-plane are given 
in Fig.~3. 
The result for the Holstein polaron indicates that the semiclassical 
criterion for the small-polaron formation corresponds to the 
condition $|T/I| \! \approx \! 0.5$. 
More or less the same result is obtained for the JT polaron for which 
Eq.~(\ref{small-polaron}) is well represented by the condition 
$|T/I| \! \approx \! 0.6$. 
In either way, we can conclude that the large-to-small polaron 
crossover occurs at around $|T/I| \! \approx \! 0.5-0.6$ and that 
a small polaron is much harder to realize in the JT system than 
the Holstein one. 

\begin{figure}[h]
\centerline{\epsfxsize=2.55truein \epsfbox{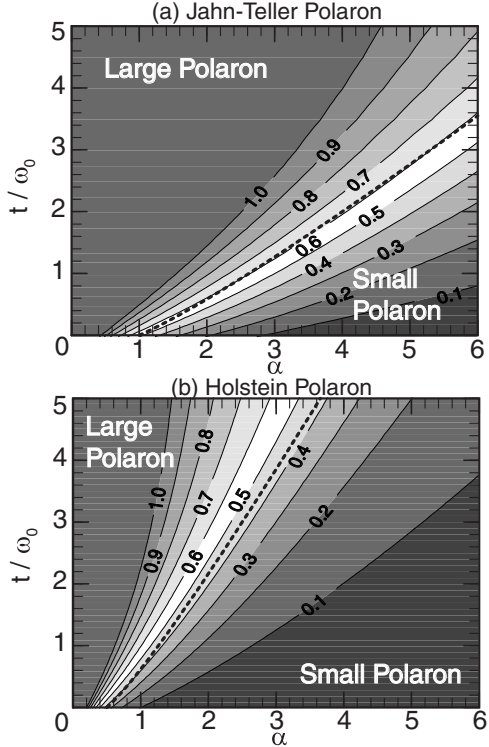} }
\caption{Contour plot for $|T/I|$ for (a) JT and (b) Holstein 
polarons. 
(Only the curves in the range $0.1-1.0$ are shown to avoid too 
many curves.)
The thick dotted curves correspond to the semiclassical criteria to 
divide large and small polarons, Eqs.~(\ref{small-polaron}) 
and (\ref{small-polaron2}).} 
\end{figure}

Three comments are in order: 
(1) In the manganese oxides, the parameters are estimated as 
$t \! \approx \! 0.2$eV, $\omega_0\! \approx \! 0.08$eV, 
and $E_{\rm JT} \! \approx \! 0.2-0.4$eV, leading to $\tilde{t} 
\! \approx \! 2.5$ and $\alpha \! \approx \! 2.5-5$, which 
covers the crossover region according to Fig.~3(a). 
This is convenient to explain the observed CMR behavior. 
(2) The very large $m^*$ in the Holstein model is unfavorable 
for the bipolaron scenario for high-$T_c$ superconductivity. 
\cite{bipolaron} 
In this respect, a smaller $m^*$ was suggested for the Fr\"ohlich 
polaron. \cite{bipolaron2} 
The same may be claimed for the JT polaron. 
(3) The electron-phonon coupling constant in $H_{\rm H}$ 
[Eq.~(\ref{Holstein-Hamiltonian})] is so determined as to give the 
same polaron effect as the JT case in the weak-coupling region 
for the proper comparison of vertex corrections. 
In this choice, the ground-state energy for $H_{\rm H}$ at $t=0$ is 
given as $(-2\alpha\!+\!1/2)\omega_0$ which is about $-2E_{\rm JT}$ 
at $\alpha \! \gg \! 1$. 
Thus, if we make an alternative choice of the coupling constant 
as to give the same polaron stabilization energy in the 
strong-coupling limit, the difference in $m^*$ between the JT and 
Holstein models looks to be much reduced, but even in this choice, 
the JT polaron has smaller $m^*$ at least by the factor of 
$1/\sqrt{\alpha}$. 

In conclusion, we have compared the $E\otimes e$ JT polaron with the 
Holstein one by using various theoretical techniques. 
Features of these polarons are exactly the same in the weak-coupling 
region, but they are different quantitatively in other regions 
due to the symmetry of pseudospin rotation; 
the JT polaron is more mobile than the Holstein one.

The author is supported by the Grant-in-Aid for Scientific Research 
(C) from the Ministry of Education, Science, Sports, and Culture of 
Japan. 


\end{multicols}

\begin{references}
\bibitem{spin} We suppress spin indices for the JT polaron 
in order to avoid confusion between ``real'' and ``pseudo'' 
spins.
\bibitem{ion-mass} The reduced mass of ions for the vibrational 
modes is normalized to unity.
\end{references}
\end{document}